# DNA SEQUENCING AND PREDICTIONS OF THE COSMIC THEORY OF LIFE

N. Chandra Wickramasinghe

Buckingham Centre for Astrobiology, The University of Buckingham, Buckingham MK18 1EG, UK; Email: ncwick@gmail.com

**Abstract**

The theory of cometary panspermia, developed by the late Sir Fred Hoyle and the present author argues that life originated cosmically as a unique event in one of a great multitude of comets or planetary bodies in the Universe. Life on Earth did not originate here but was introduced by impacting comets, and its further evolution was driven by the subsequent acquisition of cosmically derived genes. Explicit predictions of this theory published in 1979-1981, stating how the acquisition of new genes drives evolution, are compared with recent developments in relation to horizontal gene transfer, and the role of retroviruses in evolution. Precisely-stated predictions of the theory of cometary panspermia are shown to have been verified.

1. Introduction

A scientific theory has value only if it can make testable predictions. Falsifiability, according to fulfilment or otherwise of predictions, is the essence of the process by which a convergence to an objective truth can be achieved. The cometary panspermia hypothesis developed in the period 1977-1981 by Fred Hoyle and the author challenged the prevailing dogma of an origin of life in a "warm little pond" on Earth. The enormous complexity of life at a molecular genetic level prompted us to question this hypothesis on probabilistic grounds. The greatest enigma of biology – the origin of life – is in our view more likely to be resolved if we go to comets, a connected set of "warm little poinds" numbering ~ $10^{22}$ comets in our galaxy alone. The theory of the cosmic origins of life made a prediction that the infrared and ultraviolet spectra of interstellar dust would match biologically derived material (the detritus of life), which it convincingly did (Hoyle and Wickramasinghe, 2000; Wickramasinghe, 2010). It also made the prediction that the dust from comets (hitherto thought to be made of inorganic ices) must be organic, and this prediction was also vindicated by observations of comet Halley and other comets after 1986. In 2001 stratospheric air samples collected aseptically from 41km showed evidence of microorganisms of presumed cometary origin (Wainwright et al, 2003), but the possibility of terrestrial contamination cannot be absolutely ruled out. However, a later stratospheric collection revealed three new bacterial species with exceptional UV resistance properties, including one which was named *Janibacter hoylei sp.nov.* in honour of Fred Hoyle (Shivaji et al, 2009) .

The predictions of the cosmic life theory that will be reviewed in the present article are those related to the evolution of terrestrial life that were originally discussed by Hoyle and Wickramasinghe (1979, 1981). We argued that if comets brought the first life to Earth 4 billion years ago, the process of microbial additions from comets must have continued throughout geological time, and consequently played a role in the evolution of life. Such considerations were subsequently extended to a model where genetic products of evolution on a planet like the Earth were mixed on a galactic scale with products of local evolution on other planets elsewhere. It was argued that comet impacts, such as happened at the K/T boundary 65 million years ago leading to the extinction of the dinosaurs, leads also to the inevitable splash back into space of DNA fragments that carry the products of local evolution (Wallis and Wickramasinghe, 2004; Napier, 2004; Wickramasinghe et al, 2010). Even partially inactivated bacterial and viral genes expelled in this way may be capable of transmitting useful genetic information to a recipient organism in a distant place (Wickramasinghe, 2011; Wesson, 2010). In this model similar impact episodes would happen whenever the Oort cloud of comets is disturbed by the gravitational effect of a nearby massive interstellar cloud as the solar system orbits the centre of the galaxy with a period of 240 million years. We estimate the *average* time interval between successive encounters of our solar system with massive dust clouds to be about 40 million years, so that in the lifetime of the solar system of 4500 million years over a hundred such gene distribution events would have taken place for the Earth alone. Since we cannot assume the Earth and our solar system to unique in this regard it has to be assumed that similar gene dissemination sources exist for every life bearing Earth-like planet in the galaxy. As a consequence the biosphere in which Darwinian evolution occurs is naturally extended to encompass a large fraction of the volume of the galactic disc. The stochastic nature of gene acquisition events resulting from encounters with molecular clouds leads naturally to a stochastic component in biological evolution - eg sudden jumps, as is apparently observed on the Earth.

A firm prediction of the theory was that extraterrestrial viruses and bacteria – occasionally responsible for pandemics of disease – became incorporated in the germ line of survivors and provided the main driving force for evolution (Hoyle and Wickramasinghe, 1979; 1990). Although we were criticised at the time for apparently advocating a return to a primitive superstition with regard to comets and plagues, and for challenging geocentric ideas about life, advances in biology in the new millennium have produced striking evidence in our favour. Of particular importance is the discovery of horizontal gene transfer operating across a wide range of phyla, as well as of genome sequencing including the sequencing of the human genome which will be discussed in sections 3 and 4. To begin with, we shall recap in section 2, certain key predictions of the cosmic life theory with direct quotations from publications dating back to 1979 and 1982.

## 2. Predictions from 1979 and 1982

In *Diseases from Space* (1979) pp153,154 Hoyle and Wickramasinghe wrote:

"There are four steps in the complex process whereby such a virus multiplies itself the preventing of any one of which would confer immunity on an evolved plant or animal. The virus must first have an attachment protein fitted to an attachment site on the wall of a host cell. Second, the interaction of the attachment protein to the cell wall must serve eventually to strip other viral proteins away from the genetic material of the virus, which must then be afforded naked ingress to the host cell. Third, the viral genetic material must have the ability to overwrite the normal genetic program of the cell. And fourth, after multiplying in number, the new virus particles must be able to gain egress from the host cell in order to attack new cells.

With so many opportunities to frustrate the attack of viruses, and yet with evolved life forms failing to avail themselves convincingly of these opportunities, we have - within the Earth-bound point of view - the makings of a contradiction. Immunity, such as it is, consists mostly of preventing the entry of virus particles, and then only the entry of specific viruses, not of viruses in general. It would be far more effective for host cells to develop genetically so as to prevent overwriting by viruses in general. Logically, such a process must exist, because the greater quantity of information present in the genetic material of the host must be able to overcome the combined information of many viruses. Yet such a defence is never presented, except possibly very transiently during the course of an infection, when a substance called 'interferon' may have some such effect. The more likely explanation is that host cells *deliberately* avoid the apparently best form of defence. The situation is not at all that the virus is clever but that the host appears to be incorrigibly stupid. Indeed host cells even seem to invite the invasion of viruses by deliberately providing sites to which viruses can attach themselves…"

In *Proofs that Life is Cosmic*, pp 73,74 Hoyle and Wickramasinghe (1982) wrote:
"Some commentators have claimed that pathogenic viruses cannot be incident from space, for an imagined reason which they believe overrides the many facts which prove otherwise. The argument seems on minimal thought to have the attractive quality of a one-line disproof. Viruses are specific to the cells they attack it is said, as if to claim that human viruses are specific to human cells. While a minority of human viruses might be said to be specific to the cells of primates, most human viruses can actually be replicated in tissue cell cultures taken from a wide spectrum of animals, some indeed outside the mammals entirely. The proper statement therefore is that viruses are generally specific to the cells they attack to within about 150 million years of evolutionary history…….

If we had knowledge that evolution was an entirely terrestrial affair then of course it would be hard to see how viruses from outside the Earth could interact in an intimate way with terrestrially-evolved calls, but we have no such knowledge, and in the absence of knowledge all one can say is that viruses and evolution must go together. If viruses are incident from space then evolution must also be driven from space. How can this happen? Viruses do not always attack the cells they enter. Instead of taking over the genetic apparatus of the cell in order to replicate themselves, a viral particle may add itself placidly to one or other of the chromosomes. If this should happen for the sex cells of a species, mating between similarly infected individuals leads to a new genotype in their offspring, since the genes derived from the virus are copied together with the other genes whenever there is cell division during the growth of the offspring…….

A gene that happens to be useful to the adaptation of one life-form may be useless to another. Incidence from space knows nothing of such a difference, however, the gene being as likely to be added to the one form as the other. So genes that become functional in some species may exist only as nonsense genes in other species. This again is true. Genes that are useful to some species are found as redundant genes in other species. Suppose a new gene or genes to become added to the genotype (genome) of a number of members of some species. Suppose also that one or more of the genes could yield a protein or proteins that would be helpful to the adaptation of the species. The cells of those members of the species possessing the favourable new genes operate, however, in accordance with the previously existing genes, a problem arises as to how the new genes are to be switched into operation so as to become helpful to the species…..As potentially favourable genes pile up more and more, a species acquires a growing potential for large advantageous change, it acquires the potential for a major evolutionary leap, thereby punctuating its otherwise continuing state of little change – it's 'equilibrium'……."

   3. **Horizontal Gene Transfer (HGT)**

The cosmic theory of life requires that genes which are the products of evolution in a distant cosmic location (comets or planets) can, on occasion, be transferred to evolving lifeforms on the Earth (Hoyle and Wickramasinghe, 1979, 1982).   In this way evolutionary advantage or novelty could be acquired by terrestrial organisms on a stochastic basis whenever alien genetic material carrying new information is introduced to the Earth.  We thus proposed an astronomical process of horizontal gene transfers – transfer of genetic information across normal mating barriers on a cosmological scale – before it was demonstrated to operate as a process within terrestrial biology.

The mechanism of horizontal gene transfer (HGT) has now been amply documented at any rate on a terrestrial scale (Keeling and Palmer, 2008).   Boto (2010) has reviewed the role of such horizontal gene transfers in the evolution of prokaryotes and well as unicellular eukaryotes.  In prokaryotes (*E. coli*) the transfer of genes via a plasmid vector was recently demonstrated *in vivo* in the laboratory (Babic et el 2008).   There is now compelling evidence to support the once contentious view that HGT provides an important source of new genes and functions to recipient organisms and a driving force for evolution.   The operation of HGT has significantly foiled attempts to reconstruct ancient phylogenetic relationships in search of a Last Universal Common Ancestor (LUCA) (Doolittle, 1999; Jain et al, 2003).   Fig. 1 shows the tangled connections emerging from phylogenetic analysis near the base of the "tree of life", suggesting that there is no true tree of life defined by species, only a tangled web of gene connections  (Doolittle, 1999).  This clearly shows that there was no Last Universal Common Ancestor (LUCA) on the Earth, but only a universal ensemble of genes.

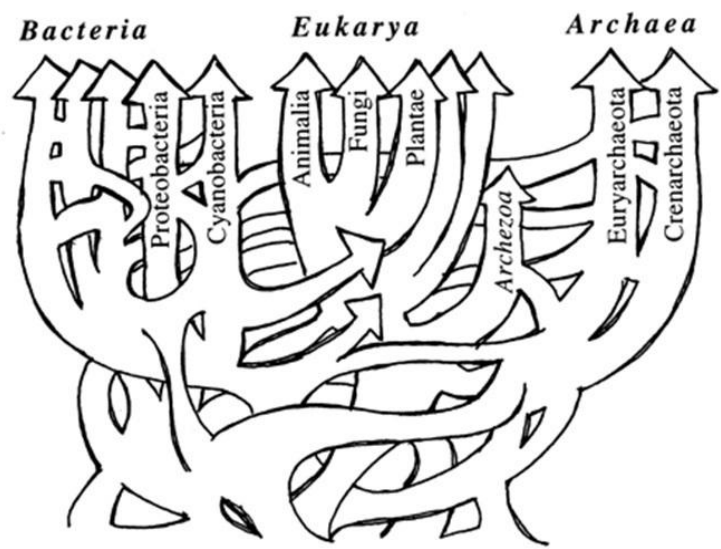

Figure 1.  Gene sequence studies reveal a tangled web of life, no tree of life, nor a convergence to a Last Universal Common Ancestor on Earth (adapted from Doolittle, 1999)

From all the available data we might infer that sudden shifts in evolution, the emergence of new traits and even the arrival of new species, occurs through HGT rather than by the slow neo-Darwinian process of mutations and natural selection (Keeling and Palmer, 2008).  Although the occurrence of neo-Darwinian gradualist evolution is not denied, it would probably be dwarfed by HGT in the long term. The phenomenon described as "punctuated equilibrium", where long periods of evolutionary stagnation are punctuated by sharp episodes of innovation and progress, is consistent with cosmically mediated HGT. The long periods of slow evolution are due to Earth-bound neo-Darwinian processes.

The successful transfer of genetic information from one organism to another in a manner that permits transmission in a host's germ line requires a vector. The vector could take the form of a plasmid, virus or a bacterium, and the host and vector is required to enter into some form of symbiosis. Eukaryotes possessing mitochondria or chloroplasts provide living evidence of horizontal gene transfers that happened in the past, with mitochondria and chloroplasts being prokaryotic endosymbionts (Keeling and Palmer, 2008). In the next section we shall cite evidence from human genome studies that similar symbiotic accommodation of viral or bacterial genes have occurred repeatedly in the primate ancestral line that led eventually to *homo sapiens.*

We conclude this section by reiterating that the evidence that now exists for purely Earth-based gene transfers can be extended to include transfers over a galactic scale, if the biosphere within which life originates and evolves is regarded as having cosmic dimensions. Transfers of alien genes would take place whenever the solar system (and Earth) encounters genetic material (viruses and bacteria) from comets or planetary systems from which such material had been expelled. The web of life displayed in the phylogenetic map of Fig. 1 is then the product of local as well as galactic "horizontal gene transfer". Dynamical routes for such transfers on a galactic scale have been identified and it is to be expected that they occurred on a stochastic basis throughout the past 4 billion years (Wallis and Wickramasinghe, 2004; Napier, 2004, Wickramasinghe, Wickramasinghe and Napier, 2010).

### 4. Viral Sequences in Genomes

Sequencing the human genome has been one of the most important scientific achievements in the new millennium. It has led to a wide range of discoveries in fundamental biology as well as medical science that are transforming our ideas about viruses, disease and evolution (Venter et al, 2001). One surprise to follow was that the number of sequences coding for proteins (genes) is as small as 20,000-25,000 rather than over 100,000 as had hitherto been suspected. The other surprise was that 8% of our genomes consist of endogenous retroviruses - RNA viruses that reverse transcribe their RNA into DNA. Their significance in causing disease as well as in contributing to evolution is only just beginning to be understood, and many astounding correspondences with statements quoted in Section 2 cannot be overlooked.

New evidence from genome studies points to frequent episodes of retroviral infections (of which HIV is an example) in almost all mammalian species. De Groot et al (2002) have identified an entire repertoire of genes (MHC class 1 genes) in chimpanzees that confer immunity against chimpanzee-derived simian immune deficiency virus. The inference is that modern chimp populations represent descendents of the survivors of a HIV-like pandemic that nearly culled the entire ancestral chimp populations in the distant past.

Following the integration of an offending retroviral gene sequence into a host's DNA, random mutations as well as development of host immunity within a few generations leads to the cessation of infectivity. Survivors of major pandemics thus carry DNA of retroviral origin bearing testimony to a history of prior infections. Such endogenisation of retroviruses often leads multiple inserts of the same viral sequence across the entire length of the host genome. The viral genes thereafter add to the potential for evolution in the long term through the action of point mutations and natural selection.

Studies of Horie et al (2010) have shown that the process of endogenisation of viruses is not confined to retroviruses. A non-retroviral RNA appears to have been incorporated in the germ line of several mammalian species, including rodents around 40 million years ago (Horie et al, 2010). With nearly half the human genome comprised of viral sequences, it would appear reasonable to suppose that evolution of hominids leading to *homo sapiens* involved a succession of viral or bacterial infections followed by endogenisation and incorporation of novel gene sequences in the evolving germ line.

In a recent paper Wang et al (2012) have shown that two immunomodulatory genes (SIGLEC) are inactive in humans, but not in related primates. It is conjectured that these genes when they were fully active could have been targets for a lethal bacterial infection that nearly culled the human population in the past, perhaps 100,000 years ago. Modern humans, on this conjecture, are descendants from the handful of survivors from a lethal bacterial pandemic.

The stochastic nature of evolution - long periods of quiescence punctuated by sudden bursts – as is evident in the fossil record is consistent with the introduction of transformative genes on occasions when the solar system came within reach of other life-bearing planetary systems and sources of novel genes. Such a picture is consistent with the prediction of the cosmic evolution model as illustrated in the quotes of section 2.

5. **Concluding remarks**

Many biologists still tend to dismiss the theory of cometary panspermia as a "controversial idea" necessarily be lacking in supportive evidence. As we have seen, this is far from being the case. One might as well ask the question: what evidence is there for the primordial soup theory and the idea that life originated on the Earth? There is certainly no direct evidence in its favour, and this theory depends for its security on an obsolete pre-Copernican philosophy. In contrast, evidence from astronomy, geology and biology continue to provide strong support for the theory of cometary panspermia (Wickramasinghe, 2010); and it should be stressed that much of this evidence was obtained *long after* our theory was formulated, as we saw in the present chapter.

In accordance with the scientific methodology pioneered by philosophers in the 17th century we can use the feedback loop of Fig. 2 to generate cycles of prediction–verification– re-affirmation to put our theory to ever more stringent tests. Needless to say there has been a veritable list of successes and confirmations over the past three decades. The loop of Fig. 2 has been enormously strengthened and amplified in recent years. The spectroscopic identification of interstellar dust and molecules in space as representing the detritus of biology, which was our starting point in the 1970s, has come into much sharper focus (Wickramasinghe, 2010). The biochemical relevance of interstellar material is now widely conceded, although a fashion remains to assert without proof that we are witnessing the operation of "prebiotic chemical evolution" on a cosmic scale. If biological evolution and replication are regarded as the only reliable facts, life always generates new life, and this must surely be so even on a cosmic scale. Prebiology, whether galactic or planetary, remains an unproven hypothesis that fails any test based upon a loop like Fig. 2. Prebiotic evolution implying a ready transition to life everywhere in the galaxy is in the author's view a mistaken remit of modern astrobiology and is likely to lead us astray. This is becoming ever more pertinent with the discovery of habitable exoplanets such as Kepler 22b, where the presence of biochemicals, when found, should properly be interpreted as evidence of extraterrestrial biology rather than prebiology.

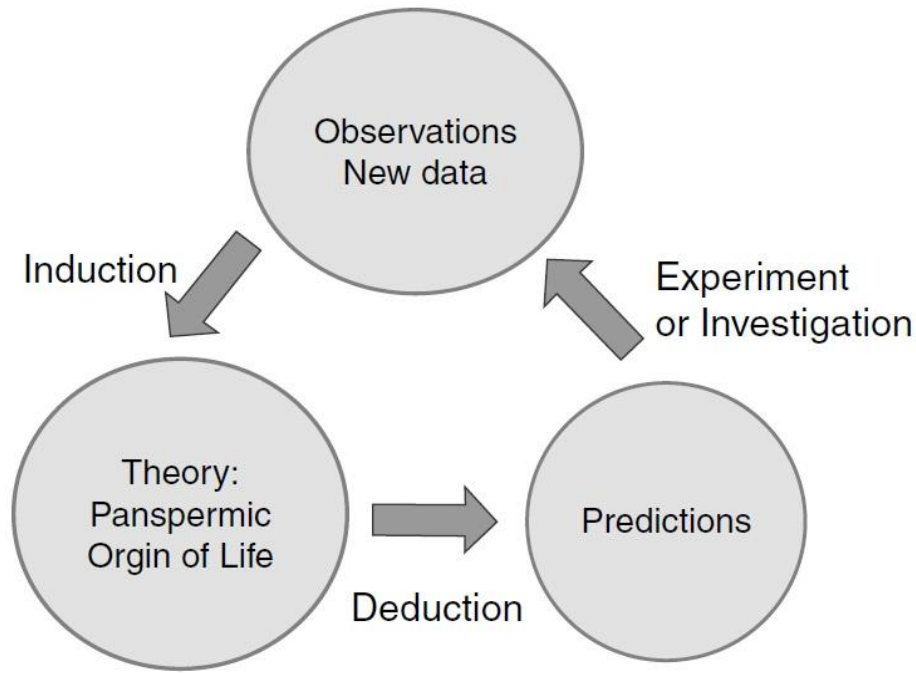

Figure 2 The deductive/inductive method of science applied to panspermia theory

In this review we pointed to the correspondences between many predictions of a model of cosmic biology, proposed over 4 decades ago, and modern developments of gene sequencing, including the human genome project. Normal conventions in science demand that such well-documented predictions as are cited in Section 2, when they come to be verified subsequently, will be acknowledged as vindication of the theory. But in the present instance the connection is not even mentioned in passing. Such is a situation is, of course, not unexpected for ideas that come under the category of Transformative Research as defined by Trevors et al (2012).

As long as the paradigm shift away from Earth-centred-origin-of-life theories remains culturally unacceptable the full significance of horizontal gene transfer and the prevalence of viral genes in our genomes will remain a puzzle. Progress towards acknowledging our cosmic ancestry will be impeded.